\documentstyle[12pt,epsfig]{article}
\begin{document}

\title{The Pure Leptonic Decays of $D_s$ Meson
 and Their Radiative Corrections}
 
\author{Guo-Li Wang$^{a,b,d}$, Chao-Hsi Chang$^{b,d}$ and Tai-Fu Feng$^{b,c,d}$}
\maketitle
\begin{center}
$a$, Department of Physics, FuJian Normal University,
 FuZhou 350007, China\\
$b$, Institute of Theoretical Physics, Academia Sinica,
 P.O.Box 2735, BeiJing 100080,China\\
$c$, Department of Physics, NanKai University,
 TianJin 300070, China\\
$d$, CCAST (World Laboratory) P.O. Box 8730, BeiJing 100080, China.

 \end{center}
\begin{center}
\begin{abstract}
The radiative corrections to the pure 
leptonic decay $D_s{\longrightarrow} 
{\ell}{ {\nu}}_{\ell}$ up-to one-loop order 
is presented.  We find the virtual photon
 loop corrections to $D_s{\longrightarrow} 
{\tau}{ {\nu}}_{\tau}$ is negative and the corresponding
branching ratio is larger than $3.51\times 10^{-3}$. Considering the
possible experimental resolutions,
 our prediction of the
radiative decay $D_s{\longrightarrow} 
{\tau}{ {\nu}}_{\tau}\gamma$ is not so large as others, and
the best channel to determine the $V_{cs}$ or $f_{D_s}$ is
 $D_s{\longrightarrow} 
{\mu}{{\nu}}_{\mu}{\gamma}$.
How to cancel the
infrared divergences appearing in the loop calculations, and
the radiative decay $D_s{\longrightarrow} 
{\ell}{{\nu}}_{\ell}{\gamma}$ is shown precisely. 
It is emphasized that the radiative decay may be
separated properly and may compare with measurements directly
as long as the theoretical `softness' of the photon corresponds to 
the experimental resolutions.

\end{abstract}
\end{center}

\section{Introduction}

\indent
   
The pure leptonic decays of the heavy-light 
meson $D_s$ are very interesting.
In principle, the pure-leptonic 
decay $D_s{\longrightarrow} {\ell}{{\nu}_{\ell}}$(see, Fig.1) 
can be used to determine the decay constant $f_{D_s}$
if the fundamental Cabibbo-Kobayashi-Maskawa matrix element $V_{cs}$ of 
Standard Model (SM) is known. Conversely if we know the value of decay
constant $f_{D_s}$ from other method\cite{as,cr,cw}, these process also can 
be used to extract the matrix element $V_{cs}$. But there are the well 
known effect of helicity suppression and there are only
one detected finial state, the measurement of pure leptonic
decays of such pseudoscalar meson is very difficult.

The pure leptonic weak decays of the pseudoscalar meson corresponding to
Fig.1 are helicity suppressed by factor of $m_{\ell}^{2}/m_{D_s}^{2}$:
\begin{equation}
\Gamma(D_s{\longrightarrow}{\ell}{\overline{\nu}_{\ell}})=
{\frac{G_{F}^{2}}{8\pi}}|V_{cs}|^{2}f_{D_s}^{2}m_{D_s}^{3}{\frac
{{m_{\ell}}^{2}}{{m_{D_s}}^{2}}}
\left(1-{\frac{{m_{\ell}}^{2}}{{m_{D_s}}^{2}}}\right)^{2},
\end{equation}
Of them only the process $D_s{\longrightarrow} {{\tau}}{\nu}_{\tau}$ 
is special, it does not suffer so much 
from the helicity suppression, and its branching ratio may reach
to $4.5\%$. However the 
produced $\tau$ will decay promptly and one more neutrino  
is generated in the cascade decay at least, thus it makes 
the decay channel difficult to be observed.
 For the channels when the lepton is $e$ or $\mu$ and their neutrinos, 
 besides the small branching ratios, there are only
one detected finial state, the measurement of
 such channels are very difficult.

Fortunately, having an extra real photon emitted in the leptonic decays, 
the radiative pure leptonic decays can escape from 
the suppression\cite{goldman,eilam,geng}, 
furthermore, as pointed out in Ref.\cite{wang}, with the extra photon 
to identify the decaying pseudoscalar meson $D_s$
in experiment from the backgrounds has advantages, since one more
particle can be detected in the detector.
Although the radiative corrections are suppressed by an
extra electromagnetic coupling constant $\alpha$, it will not be
suppressed by the helicity suppression. Therefore, the
radiative decay may be comparable, even larger than
the corresponding pure leptonic decays\cite{goldman,eilam,geng}.
So, the problem to increase   
the accuracy of the theoretical calculation, at least up to the first 
order radiative corrections, emerges.

The radiative pure leptonic decays, theoretically, have infrared 
divergences and will be canceled with those from loop corrections of 
the pure leptonic decays. In all the existing calculation of 
radiative decays\cite{goldman,eilam,geng}, this part is ignored, since they do not include
the radiative corrections of the pure leptonic decays. 
In this paper, we are interested in 
considering the radiative decays and the pure leptonic decays with
one-loop radiative corrections together. Since the process
$D_s{\longrightarrow} {{\tau}}{\nu}_{\tau}$ dose not suffer the 
helicity supression and has a large branching ratio, the corresponding
loop correction(virtual photon) to these process should
 has a considerable larger branching
ratio, at least comparing with the radiative decay, and can not be ignored. 

The paper is organized as follow: in Section II, we present
the calculations of radiative pure leptonic contributions:
{\it i)} besides the leptonic decays shown as Fig.1,
the radiative decays into three family leptons with a real photon 
appearing in the final state as shown in Fig.2; {\it ii)} the virtual photon
corrections to the pure leptonic decays, 
i.e., a photon in loop as shown in Figs.3.1, 3.2, 3.3, 3.4.  
In Section III, we evaluate the values of the decays. 
As uncertainties the dependence of the results on 
the parameters appearing in the considered model is
discussed. Finally, preliminary conclusion is obtained.

%\par
\section{The Radiative Leptonic Contributions}

%\indent

Let us start with the radiative decay. These contributions are 
corresponding to the four diagrams 
in Fig.2. According to the constituent quark model which is formulated 
by Bethe-Salpeter (B.-S.) equation, the amplitude turns out to be
the four terms $M_i (i=1,2,3,4)$:
$$M_1=Tr \left[\int\frac{d^{4}q}{(2\pi)^{4}}\chi(p, q) i 
\left(\frac{G_{F}m_{w}^{2}}{\sqrt{2}
}\right)^{\frac{1}{2}}{\gamma}_{\mu}(1-\gamma_{5})V_{cs}\right]\times $$
$${\frac{ i \left(-g^{\mu\nu}+
\frac{p^{\mu}p^{\nu}}{m_{w}^{2}}\right)}{p^{2}-m_{w}^{2}}}
  i e [(p'+p)_{\lambda}g_{\nu\rho}+
(k-p')_{\nu}g_{\rho\lambda}+(-p-k)_{\rho}g_{\nu\lambda} ] 
{\epsilon}^{\lambda}\times$$

\begin{equation}
\frac{ i \left(-g^{{\rho}{\sigma}}+
\frac{(p-k)^{\rho}(p-k)^{\sigma}}{m_{w}^{2}}\right)}{(p-k)^{2}-
m_{w}^{2}}\overline { \ell }\frac{ig}{2\sqrt{2}}\gamma_{\sigma}
(1-{\gamma}_5)\nu_\ell,
\end{equation}

$$M_2=Tr \left [ \int\frac{d^{4}q}{(2\pi)^{4}}\chi(p, q) i 
\left (\frac{G_{F}m_{w}^{2}}{\sqrt{2}
}\right )^{\frac{1}{2}}{\gamma}_{\mu}(1-\gamma_{5})V_{cs}
\right ]{\frac{ i \left(-g^{\mu\nu}+
\frac{p^{\mu}p^{\nu}}{m_{w}^{2}}\right)}{p^{2}-m_{w}^{2}}}$$

\begin{equation}
 \times\overline { \ell }(-ie)\not\! { \epsilon } \frac{i}{\not\! 
 {k}_{\ell}-m_{\ell}}\frac{ig}{2\sqrt{2}}\gamma_{\nu}(1-{\gamma}_5)\nu_\ell,
\end{equation}
 
$$M_3=Tr \left [ \int\frac{d^{4}q}{(2\pi)^{4}}\chi(p, q) 
i\left (\frac{G_{F}m_{w}^{2}}{\sqrt{2}
}\right )^{\frac{1}{2}}{\gamma}_{\mu}(1-\gamma_{5})V_{cs}
\frac{i}{\frac{m_{s}}{m_{s}+m_{c}}\not\! {p}
+\not\! {q}-\not\! {k}-m_{s}}\left(-i \frac{e}{3}\not\! 
{\epsilon}\right)\right ] $$

\begin{equation}
\times\frac{ i \left(-g^{{\mu}{\sigma}}+
\frac{(p-k)^{\mu}(p-k)^{\sigma}}{m_{w}^{2}}\right)}{(p-k)^{2}-
m_{w}^{2}}\overline { \ell }\frac{ig}{2\sqrt{2}}
\gamma_{\sigma}(1-{\gamma}_5)\nu_\ell,
\end{equation}

$$M_4=Tr \left [ \int\frac{d^{4}q}{(2\pi)^{4}}\chi(p, q)
\left( i \frac{2e}{3}\not\! {\epsilon}\right)
\frac{i}{-(\frac{m_{c}}{m_{s}+m_{c}}\not\! {p}
+\not\! {q}-\not\! {k})-m_{c}} i \left (\frac{G_{F}m_{w}^{2}}{\sqrt{2}
}\right )^{\frac{1}{2}}{\gamma}_{\mu}(1-\gamma_{5})V_{cs}\right] $$

\begin{equation}
\times\frac{ i \left(-g^{{\mu}{\sigma}}+
\frac{(p-k)^{\mu}(p-k)^{\sigma}}{m_{w}^{2}}\right)}{(p-k)^{2}-
m_{w}^{2}}\overline { \ell } \frac{ig}{2\sqrt{2}}\gamma_{\sigma}
(1-{\gamma}_5)\nu_\ell,
\end{equation}
where $\chi(p, q)$ is Bethe-Salpeter wave function of the meson $D_s$; 
$p$ is the momentum of $D_s$; ${\epsilon}, k$ are the polarization 
vector and momentum of the emitted photon.   
In the quark model, the momenta of $\bar {s}, c$-quarks
inside the bound state, i.e., the $D_s$ meson, are:
%\begin{center}
$$\displaystyle p_{s}={\frac{m_{s}} {m_{s}+m_{c}}}p+q; \;\;\; 
p_{c}={\frac{m_{c}} {m_{s}+m_{c}}}p-q ,$$
%\end{center}
where $q$ is the relative momentum of the two quarks
inside the $D_s$ meson. As the $D_s$ meson is a nonrelativistic
bound state in nature, so the higher order relativistic corrections
may be computed precisely, but being an approximation for a $S$-wave
state, and focusing the light on the radiative decay corrections only
at this moment now, we ignore $q$ dependence, i.e., we may
still have the non-relativistic 
spin structure for the wave function of the meson $D_s$ (a $^1S_0$ state)
correspondingly:
$$\int\frac{d^{4}q}{(2\pi)^{4}}\chi(p, q)=
\frac{\gamma_{5}({/}\!\!\! {p}{+m})}{2\sqrt{m}}\psi(0).$$ 
Here $\psi(0)$ is the wave function at origin, and by definitions
it connects to the decay constant $f_{D_s}$:
$$f_{D_s}=\frac{\psi(0)}{2\sqrt {m}},$$
where m is the mass of $D_s$ meson. Moreover we note that for 
convenience we take unitary gauge for weak bosons to do the calculations
throughout the paper.
With a straightforward computation, the amplitude can be simplified as:   
$$M_{1}={\frac{-4Ai}{(m^{2}-m_{w}^{2})(m^{2}-2p\cdot k-m_{w}^{2})}}\times$$
\begin{equation}
\overline {\ell}\left (-1+\frac{m^{2}}{m_{w}^{2}}\right )
\left[ p\cdot\epsilon(\not\! {k}-\not\! {p})-(2p\cdot{k}-m^{2})
\not\! {\epsilon} \right](1-\gamma_{5})\nu_{\ell},
\end{equation}

\begin{equation}
M_{2}={\frac{4Ai}{\left(m^{2}-m_{w}^{2}\right)(2k_{1}\cdot{k})}}
\overline {\ell}\left(-1+\frac{m^{2}}{m_{w}^{2}}\right)
\not\! {\epsilon}(\not\! {k}_{1}+\not\! {k}+m_{e})\not\! 
{p}(1-\gamma_{5})\nu_{\ell},
\end{equation}

\begin{equation}
\begin{array}{lcr}
M_{3}+M_{4}&=&\frac{-4Ai}{(-p\cdot{k})(m^{2}-2p\cdot k-m_{w}^{2})}
\overline {\ell}
\left\{\left[-(p\cdot{\epsilon})\not\! {p}+
(p\cdot{\epsilon}\not\! {k}-p\cdot{k}\not\! 
{\epsilon})s_{2} \right.\right.\\ [2mm]
 &+& \left. \left. is_{1}\varepsilon^{\alpha\mu\beta\nu}p_{\alpha}
\epsilon_{\mu}k_{\beta}\gamma_{\nu}\right]+
\frac{(\not\! {p}-\not\! {k})}{m_{w}^{2}}p\cdot{\epsilon}(m^{2}-p\cdot{k})
\right\}(1-\gamma_{5})\nu_{\ell},
\end{array}
\end{equation}
where $k_{1}$ is the momentum of the charged lepton, and
$$A=\frac{\psi(0)\left(\frac{G_{F}m_{w}^{2}}{\sqrt{2}}\right)^{\frac{1}{2}}
V_{cs}eg}{2\sqrt{m}2\sqrt{2}}=
\frac{\psi(0)\left(\frac{G_{F}m_{w}^{2}}{\sqrt{2}}\right)V_{cs}e}{2\sqrt{m}};$$
$$s1=-\frac{m_{s}+m_{c}}{6m_{s}}+\frac{m_{s}+m_{c}}{3m_{c}}; 
\;\; s2=\frac{m_{s}+m_{c}}{6m_{s}}+\frac{m_{s}+m_{c}}{3m_{c}}.$$

As a matter of fact,
there is infrared infinity when performing phase space integral about
the square of matrix element at the soft photon limit. 
It is known that the infrared infinity can be
cancelled completely by that of the 
loop corrections to the corresponding pure leptonic decay $D_s\to \ell\nu$.

In Eqs.(7) and (8), the infrared terms can be
read out: 
\begin{equation}
M^{i}=M_{2}^{i}+M_{3}^{i}+M_{4}^{i}
=\frac{4Ai}{m_{w}^{2}}\overline{\ell}\left[\frac{k_{1}\cdot{\epsilon}\not\! 
{p}}{k_{1}\cdot{k}}-
\frac{p\cdot{\epsilon}\not\! {p}}{p\cdot{k}}\right](1-\gamma_{5})\nu_{\ell}.
\end{equation}

As the diagrams (g), (h), (i), (j) in Figs.3.3, 3.4 always
have a further suppression factor $m^{2}/{m_{w}}^{2}$
to compare with the other loop diagrams, and there is no infrared infinity in
these four loop diagrams, we may ignore 
the contributions from these four diagrams safely.
Furthermore we should note that in our 
calculations throughout the paper, the dimensional regularization 
to regularize both infrared and ultraviolet divergences is adopted, 
while the on-mass-shell renormalization for the ultraviolet divergence 
is used.

If Feynman gauge for photon is taken (we always do so in the paper), 
the amplitude corresponding to the diagrams (a), (b) of Fig.3.1 
can be written as:

$$M_{(2)}(a)={\frac{2}{3}}eA\int\frac{d^{4}l}{(2\pi)^{4}}
\left[\frac{-4i\varepsilon^{\alpha\mu\beta\nu}p_{\alpha}l_{\beta}-4(p_{\mu}l_{\nu}
-{p\cdot l}g_{\mu\nu}+p_{\nu}l_{\mu})+\frac{8m_{c}}{m_{s}+m_{c}}p_{\mu}p_{\nu}}
{l^{2}(l^{2}-2p\cdot l-m_{w}^{2})(l^{2}-{\frac{2m_{c}}{m_{s}+m_{c}}}p\cdot {l}
)[l^{2}-2l\cdot {(p-k_{2})}]}\right]$$
\begin{equation}
\times\overline{\ell}[2(p-k_{2})_{\mu}-\gamma_{\mu}{\not\!   {l}}]
(-\gamma_{\nu})(1-\gamma_{5})\nu_{\ell},
\end{equation}

$$M_{(2)}(b)={-\frac{1}{3}}eA\int\frac{d^{4}l}{(2\pi)^{4}}
\left[\frac{-4i\varepsilon^{\alpha\mu\beta\nu}p_{\alpha}l_{\beta}+4(p_{\mu}l_{\nu}
-{p\cdot l}g_{\mu\nu}+p_{\nu}l_{\mu})-\frac{8m_{s}}{m_{s}+m_{c}}p_{\mu}p_{\nu}}
{l^{2}(l^{2}-2p\cdot l-m_{w}^{2})(l^{2}-{\frac{2m_{s}}{m_{s}+m_{c}}}p\cdot {l}
)[l^{2}-2l\cdot {(p-k_{2})}]}\right]$$
\begin{equation}
\times\overline{\ell}[2(p-k_{2})_{\mu}-\gamma_{\mu}{\not\! {l}}]
(-\gamma_{\nu})(1-\gamma_{5})\nu_{\ell},
\end{equation}
where the $l$, $k_{2}$ denote the momenta of the loop and the
neutrino respectively. These two terms 
also have infrared infinity when integrating out the loop momentum $l$.
 
After doing the on-mass-shell subtraction, the terms 
corresponding to vertex and 
self-energy diagrams (c), (d), (e), (f) can be written as:
%\begin{center}\begin{eqnarray}
$$M_{(2)}(c+d+e+f)=\frac{ieA}{4\pi^{2}}\overline{\ell}{\not\! p}
(1-\gamma_{5})\nu_{\ell}\times\left[
 ln(4)-\frac{8}{9}+\frac{2}{9}{\frac{m_{s}-m_{c}}{m_{s}+m_{c}}}
ln\left(\frac{m_{s}}{m_{c}}\right)\right.$$%\nonumber\\
 $$+\left(\frac{2}{9}+\frac{8}{9}{\frac{m_{c}}{m_{s}+m_{c}}}\right)
ln\left(\frac{m_{s}+m_{c}}{m_{s}}\right)
 +\left(\frac{8}{9}+\frac{8}{9}{\frac{m_{c}}{m_{s}+m_{c}}}
\right)ln\left(\frac{m_{s}+m_{c}}{m_{c}}\right)$$%\nonumber\\
\begin{equation} +\left.\frac{2}{\varepsilon_{I}}-2\gamma+
ln\left(\frac{4\pi\mu^{2}}{m^{2}}\right)+
 ln\left(\frac{4\pi\mu^{2}}{m_{e}^{2}}\right)\right].
\end{equation}
%\end{eqnarray} \end{center}

Now let us see the cancellation of the infrared divergencies
precisely. The infrared parts of the decay widths
which are from the interference of
the self-energy and vertex correction diagrams
with the tree diagrams:
$$\delta\Gamma_{s,v}^{infrared}=
\displaystyle\left(\frac{\alpha
V_{cs}^{2}f^2_{D_s}G^{2}_{F}Mm^{2}_{\ell}}{16{\pi}^2}\right)\left[
-\frac{34}{9}-\frac{4}{9}{\frac{m_{s}-m_{c}}{m_{s}+m_{c}}}ln\left(\frac{m_{s}}
{m_{c}}\right)\right.$$
$$-\left(\frac{2}{9}+\frac{4}{9}{\frac{m_{c}}{m_{s}+m_{c}}}\right)ln\left(\frac{
m_{s}+m_{c}}{m_{s}}\right)
-\left(-\frac{4}{9}+\frac{4}{9}{\frac{m_{s}}{m_{s}+m_{c}}}\right)ln\left(\frac{m
_{b}+m_{c}}{m_{c}}\right)$$
\begin{equation}
-\left.\frac{2}{\varepsilon_{I}}+2\gamma-ln\left(\frac{4\pi\mu^{2}}{m^{2}}\right
)-
 ln\left(\frac{4\pi\mu^{2}}{m_{e}^{2}}\right)\right] \; ;
\end{equation}
the infrared part of the decay widths from the interference
of the "box" correction diagrams with the tree diagrams:
\begin{equation}
\delta\Gamma_{box}^{infrared}=
\left(\frac{\alpha V_{cs}^{2}f^2_{D_s}G^{2}_{F}Mm^{2}_{\ell}}
{16{\pi}^2}\right)\left[
-\left(\frac{1}{\varepsilon_{I}}-\gamma\right)ln\left(\frac{m^2_{\ell}}{M^2}
\right)-
ln\left(\frac{m^2_{\ell}}{M^2}\right)ln\left(\frac{4\pi\mu^2}{M^2}\right)\right]\; ;
\end{equation}
the infrared part of the decay width from the real photon emission:
$$ \delta\Gamma_{real}^{infrared}= 
\left(\frac{\alpha V_{cs}^{2}f^2_{D_s}G^{2}_{F}Mm^{2}_{\ell}}
{16{\pi}^2}\right)$$
\begin{equation}
\cdot \left[\frac{2}{\varepsilon_I}-2\gamma+\left(\frac{1}{\varepsilon_{I}}-
\gamma\right)ln\left(\frac{m^2_{\ell}}{M^2}\right)
+\left( 2+ ln\left( \frac{m^2_{\ell}}{M^2} \right)
\right) ln \left(\frac{4\pi\mu^2}{4(\Delta E)^2} \right) \right] \;.
\end{equation}
Here $\Delta{E}$ is a small energy, which corresponds to the
experimental resolution of a soft photon so that the phase space of the
emitting photon in fact is divided into a soft and a hard part
by $\Delta{E}$. Where $\mu$ is the dimensional parameter 
appearing in the dimensional regularization. 
It is easy to check that when adding up all the parts: the real photon
emission $\delta\Gamma_{real}^{infrared}$ and the virtual photon 
corrections $\delta\Gamma_{s,v}^{infrared}, \delta\Gamma_{box}^{infrared}$, 
the infrared divergences $\frac{2}{\varepsilon_{I}}-\gamma$ are canceled 
totally and the $\mu$ dependence is also cancelled. Hence we may be sure 
that we finally obtain the pure leptonic widths for the $D_s$ meson decays 
to the three families of leptons which are accurate up-to the `next'-order 
corrections and `infrared finite' but depend on the experimental resolution
$\Delta{E}$.

\section{Numerical Results and Discussion}

%\indent 
First of all, let us show the `whole' leptonic decay widths,
i.e., the sum of the corresponding radiative decay widths 
and the corresponding pure leptonic decay widths
with radiative corrections, and put them in Table (1).
Why we put the radiative decay 
and the pure leptonic decay with radiative corrections together here is to
make the width not to depend on the experimental 
resolution for a soft photon. The values for the parameters $\alpha=1/132$, 
$|V_{cs}|=0.974$\cite{groom}, $m_{D_s}=1.9686$ GeV, $m_{s}=1.5$ GeV, $m_{c}=1.7$ GeV and
$f_{D_s}=0.24$ GeV\cite{allton}.\\

\begin{center}
Table (1) The `Whole' Leptonic Decay Widths (in unit GeV)\\ 
\vspace{2mm}
\begin{tabular}{|c|c|} \hline
 $\Gamma_{e}(10^{-17})$ &   3.593  \\ \hline
 $\Gamma_{\mu}(10^{-15})$ &  6.604  \\ \hline
 $\Gamma_{\tau}(10^{-14})$ & 5.807  \\\hline
\end{tabular}
\end{center}

For comparison, the width of each pure leptonic decay 
at tree level with the same parameters as those in Table (1) is 
put in Table (2).
                               
\begin{center}
Table (2) The Pure Leptonic Decay Widths (in unit GeV) of Tree Level\\
\vspace{2mm}
\begin{tabular}{|c|c|} \hline
 $\Gamma_{e}(10^{-19}$)& 1.521 \\ \hline
 $\Gamma_{\mu}(10^{-15}$) &  6.463\\ \hline
 $\Gamma_{\tau}(10^{-14}$)&  6.30 \\ \hline
\end{tabular}
\end{center}

If the lifetime of $D_s$ meson is $\tau(D_s)=0.469\times10^{-12}s$\cite{groom}, 
the corresponding branching ratios 
are showed in Tables (3), (4).

\begin{center}
Table (3) Branching Ratios of the `Whole' Leptonic Decays \\
\vspace{2mm}
\begin{tabular}{|c|c|} \hline
 $B_{e}(10^{-5})$ & 2.56\\ \hline
 $B_{\mu}(10^{-3})$ & 4.706 \\ \hline
 $B_{\tau}(10^{-2})$& 4.138 \\ \hline
\end{tabular}
\end{center}

\begin{center}
Table (4) Tree Level Branching Ratios of The Pure Leptonic Decays\\
\vspace{2mm}
\begin{tabular}{|c|c|} \hline
 $B_{e}(10^{-7})$&1.084\\ \hline
 $B_{\mu}(10^{-3})$ &4.605\\ \hline
 $B_{\tau}(10^{-2})$&4.489\\ \hline
\end{tabular}
\end{center}

We can see that, the whole decay
widths $\Gamma_{e}$ and $\Gamma_{\mu}$ in Table (1) are 
larger than the corresponding decay widths of tree lever 
in Table (2), while the $\Gamma_{\tau}$ in Table (1)
is smaller than the one in Table (2).
This means the sums of radiative leptonic decay and loop correction as 
the first order contributions of pure leptonic decay are positive
for the decays with a $e$ or $\mu$ in the finial state, and negative
for the one with a $\tau$ in the finial state. 
It also show that the contributions
of loop corrections are negative,
the dominate contributions of first order corrections 
to the pure leptonic $D_s$ decays are radiative decays 
 when the lepton is $e$ or $\mu$,  and  
is loop corrections when the lepton is $\tau$. So, the loop contributions
are important for the decays $D_s\to \mu \nu_{\mu}$ 
and $D_s\to \tau \nu_{\tau}$, especially for the later. Through Table (3)
and (4), we obtain that the rediative decay has a branching ratio 
$Br(D_s\to \mu \nu_{\mu}\gamma)> 1.01\times 10^{-4}$ and the loop
correction to $D_s\to \tau \nu_{\tau}$ has a branching ratio 
$Br> 3.51\times 10^{-3}$.

To see the contributions of the radiative decays precisely
we present the radiative decay widths with a cut of the
photon energy, i.e., the widths of the radiative decays 
$D_s\rightarrow l\nu\gamma$ with the photon energy $E_\gamma \geq k_{min}$
as the follows: $k_{min}=0.00001$ GeV, $k_{min}=0.0001$ GeV, 
$k_{min}=0.001 $ GeV, $k_{min}=0.01 $ GeV and $k_{min}=0.1 $ GeV 
respectively in Table (5).

\begin{center}
Table (5): The Radiative Decay Widths (in unit GeV)
 with cuts of the photon momentum
 
\vspace{2mm}
\begin{tabular}{|c|c|c|c|}\hline
$k_{min}$ &$\Gamma _e $&$\Gamma _{\mu}$&$\Gamma _{\tau} $  \\ \hline
GeV  & $10^{-17}$ & $10^{-16} $& $10^{-18} $  \\ \hline
0.00001& 3.582& 6.878& 8.892  \\ \hline
0.0001&  3.581& 5.484& 6.452  \\ \hline
0.001 &  3.580& 4.091& 4.025  \\ \hline
0.01 &   3.578& 2.704& 1.708  \\ \hline
0.1&     3.474& 1.363& 1.020  \\ \hline
\end{tabular}
\end{center}
The corresponding branching ratios of the radiative decay with cuts of photon
momentum are showed in Table (6)
\begin{center}
Table (6): The Radiative Decay branching ratios 
 with cuts of the photon momentum and the results of Ref\cite{goldman,eilam,geng}\\
 
\vspace{2mm}
\begin{tabular}{|c|c|c|c|}\hline
$k_{min}$ &$Br_e $&$Br_{\mu}$&$Br_{\tau} $  \\ \hline
GeV  & $10^{-5}$ & $10^{-4} $& $10^{-6} $  \\ \hline
0.00001& 2.552& 4.901& 6.336  \\ \hline
0.0001&  2.552& 3.908& 4.597  \\ \hline
0.001 &  2.551& 2.915& 2.868  \\ \hline
0.01 &   2.549& 1.927& 1.217  \\ \hline
0.1&     2.475& 0.971& 0.727  \\ \hline
Ref\cite{goldman}&   10 &   1  &     \\ \hline
Ref\cite{eilam}  &   17 & 1.7    &     \\ \hline
Ref\cite{geng}  &   7.7 &  2.6   &   320  \\ \hline
\end{tabular}
\end{center}

Considering the possible experimental resolutions,
 our prediction of the radiative decay widths $\Gamma(D_s\to e\nu_e\gamma)$ and
$\Gamma(D_s\to \mu\nu_{\mu}\gamma)$ are close to the values in
 Ref\cite{goldman,eilam,geng},
but our prediction of $\Gamma(D_s\to \tau\nu_{\tau}\gamma)$ is much smaller than
the one in Ref\cite{geng}. In our model,
if we using a smaller cut $k_{min}$,
 the obtained
$Br(D_s\to \ell\nu_{\ell}\gamma)$ will become larger, 
 but since the decay widths depend
on $Log(k_{min})$, the change of branching ratios
 will be not so much on the selecting of $k_{min}$, we can
see this in Table (5) and (6), and for another example, if $k_{min}=1\times 10^{-10}$, 
the $Br(D_s\to e\nu_{e}\gamma)=3.59\times 10^{-5}$, 
$Br(D_s\to \mu\nu_{\mu}\gamma)=1.38\times 10^{-3}$, 
$Br(D_s\to \tau\nu_{\tau}\gamma)=2.11\times 10^{-5}$, but 
it is very difficult to obtain so small a $k_{min}$ in experiment.
 We can conclude that the best radiative decay channel is easy
 to search is $D_s\to \mu\nu_{\mu}\gamma$. 

For the convenience to compare with experiments, we present the photon
spectrum of the radiative decays in Fig.4 and Fig.5.   
In addition, we should note that the widths are quite sensitive to 
the decay constant $f_{D_s}$, and are sensitive to
 the values of the quark masses 
$m_{s}$ and $m_c$.

%\begin{}

\begin{figure}\begin{center}
   \epsfig{file=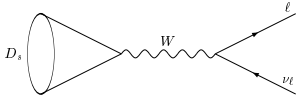, bbllx=146pt,bblly=300pt,bburx=371pt,bbury=393pt,
width=4cm,angle=0}
\caption{\bf Tree diagram for $D_s\longrightarrow\ell\nu_{\ell}$.}
%\label{fig}
\end{center}
\end{figure}

\begin{figure}\begin{center}
   \epsfig{file=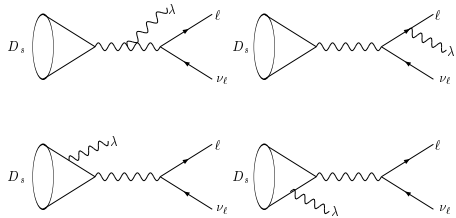, bbllx=157pt,bblly=497pt,bburx=505pt,bbury=683pt,
width=6cm,angle=0}
\caption{\bf Diagrams for $D_s\longrightarrow \ell \nu \gamma $.}
%\label{fig}
\end{center}
\end{figure}
\setcounter{figure}{2}

\begin{figure}\begin{center}
   \epsfig{file=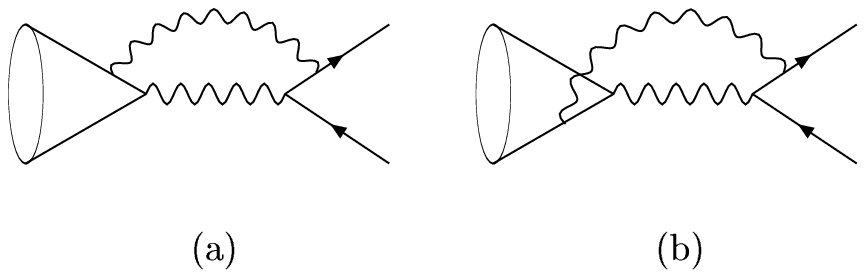, bbllx=146pt,bblly=230pt,bburx=400pt,bbury=316pt,
width=4cm,angle=0}
\caption{\bf 1. Box-loop diagrams for $D_s \longrightarrow \ell \nu$.}
%\label{fig}
\end{center}
\end{figure}

\setcounter{figure}{2}
\begin{figure}\begin{center}
   \epsfig{file=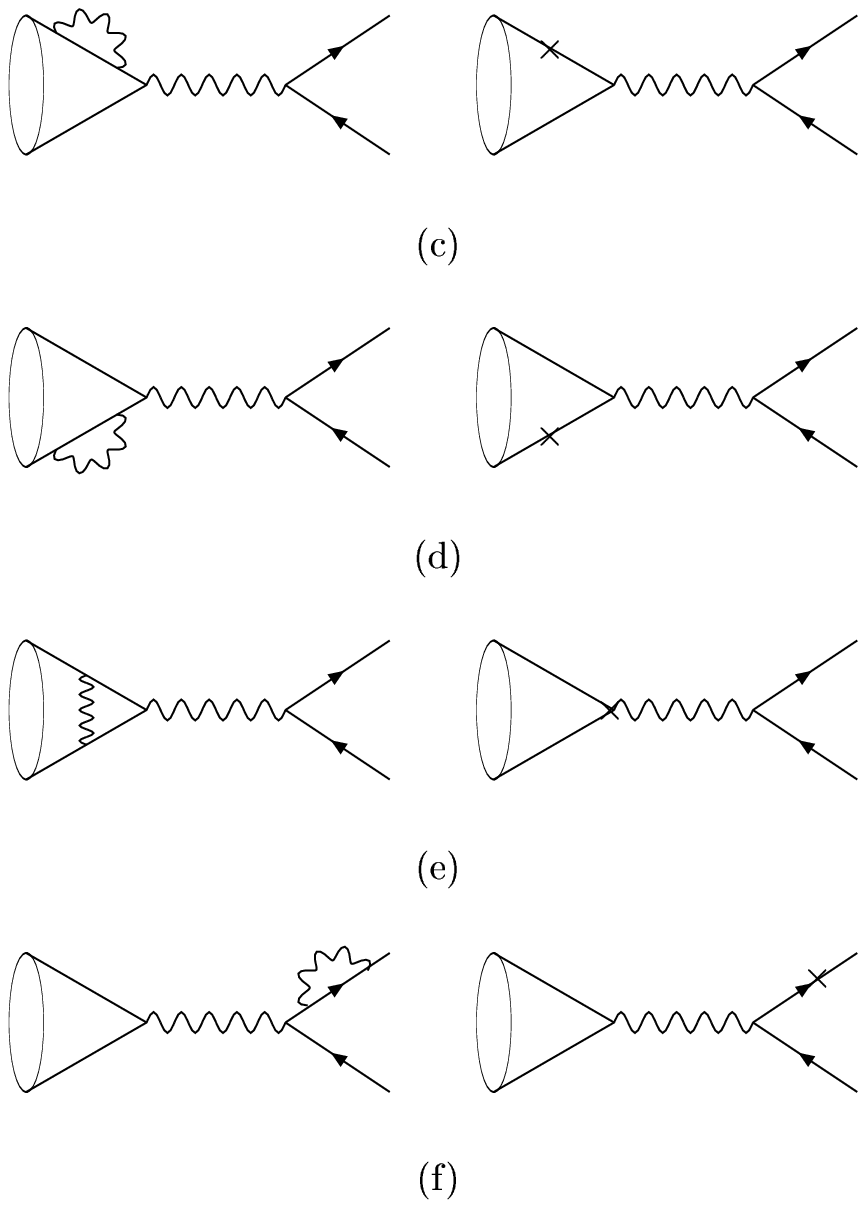, bbllx=277pt,bblly=312pt,bburx=536pt,bbury=676pt,
width=6cm,angle=0}
\caption{\bf 2. Self-energy and vertex diagrams for $D_s \longrightarrow \ell \nu$.}
%\label{fig}
\end{center}
\end{figure}

\setcounter{figure}{2}
\begin{figure}\begin{center}
\epsfig{file=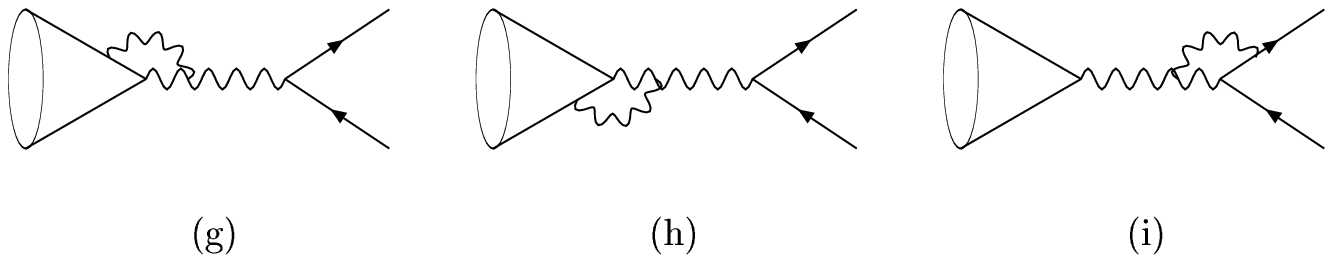, bbllx=146pt,bblly=234pt,bburx=530pt,bbury=312pt,
width=6cm,angle=0}
\caption{\bf 3. Vertex diagrams for $D_s \longrightarrow \ell \nu$.}
%\label{fig}
\end{center}
\end{figure}

\setcounter{figure}{2}
\begin{figure}\begin{center}
\epsfig{file=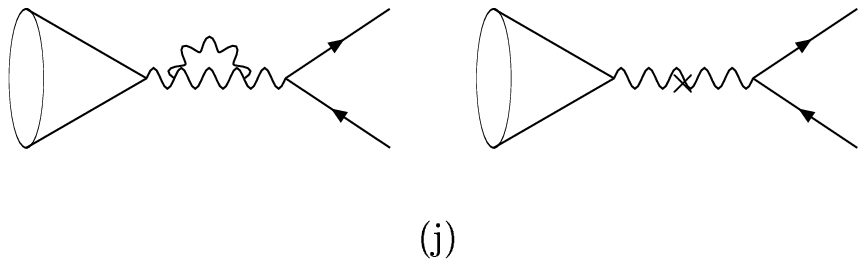, bbllx=277pt,bblly=535pt,bburx=530pt,bbury=610pt,
width=5cm,angle=0}
\caption{\bf 4. Self-energy diagrams for $D_s \longrightarrow \ell \nu$.}
%\label{fig}
\end{center}
\end{figure}

\begin{figure}\begin{center}
   \epsfig{file=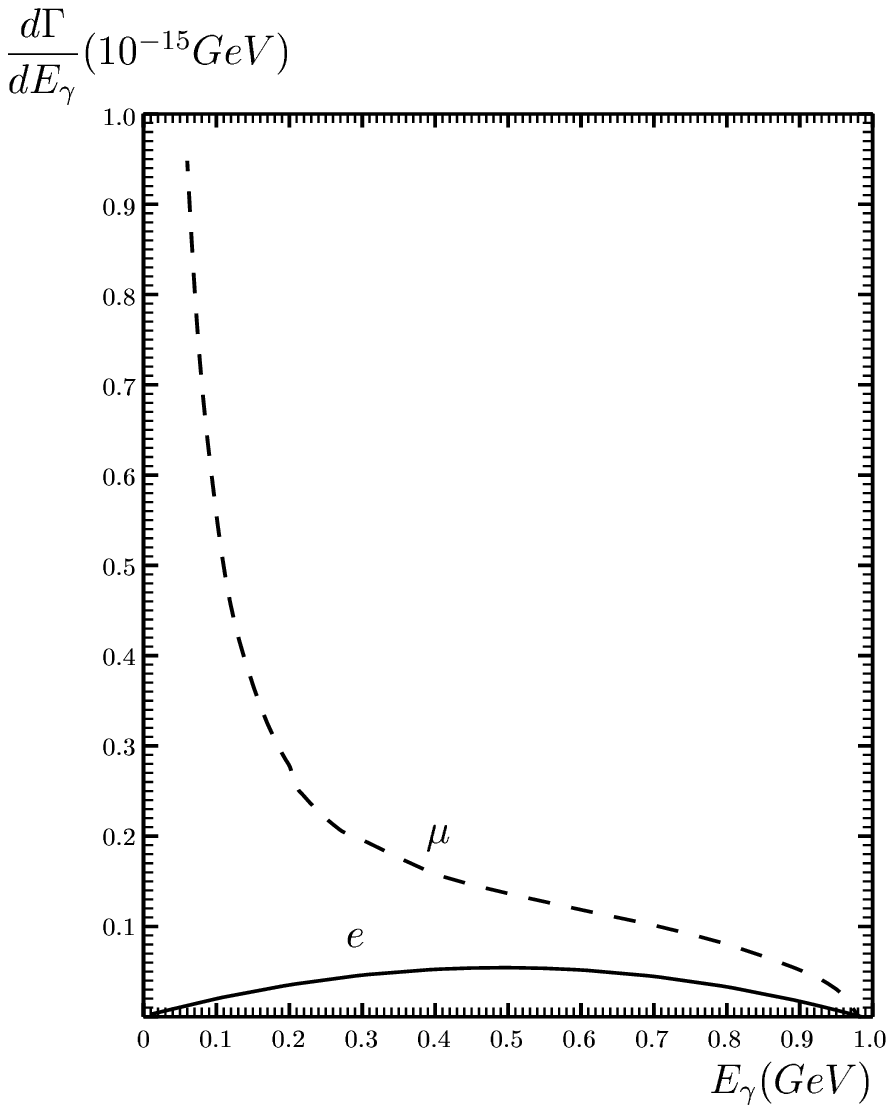, bbllx=216pt,bblly=365pt,bburx=450pt,bbury=680pt,
width=5cm,angle=0}
\caption{\bf Photon energy spectra of radiative decays  $D_s\longrightarrow\ell{\nu_{\ell}}
\gamma(\ell=e, \mu)$.}
%\label{fig}
\end{center}
\end{figure}

\begin{figure}\begin{center}
   \epsfig{file=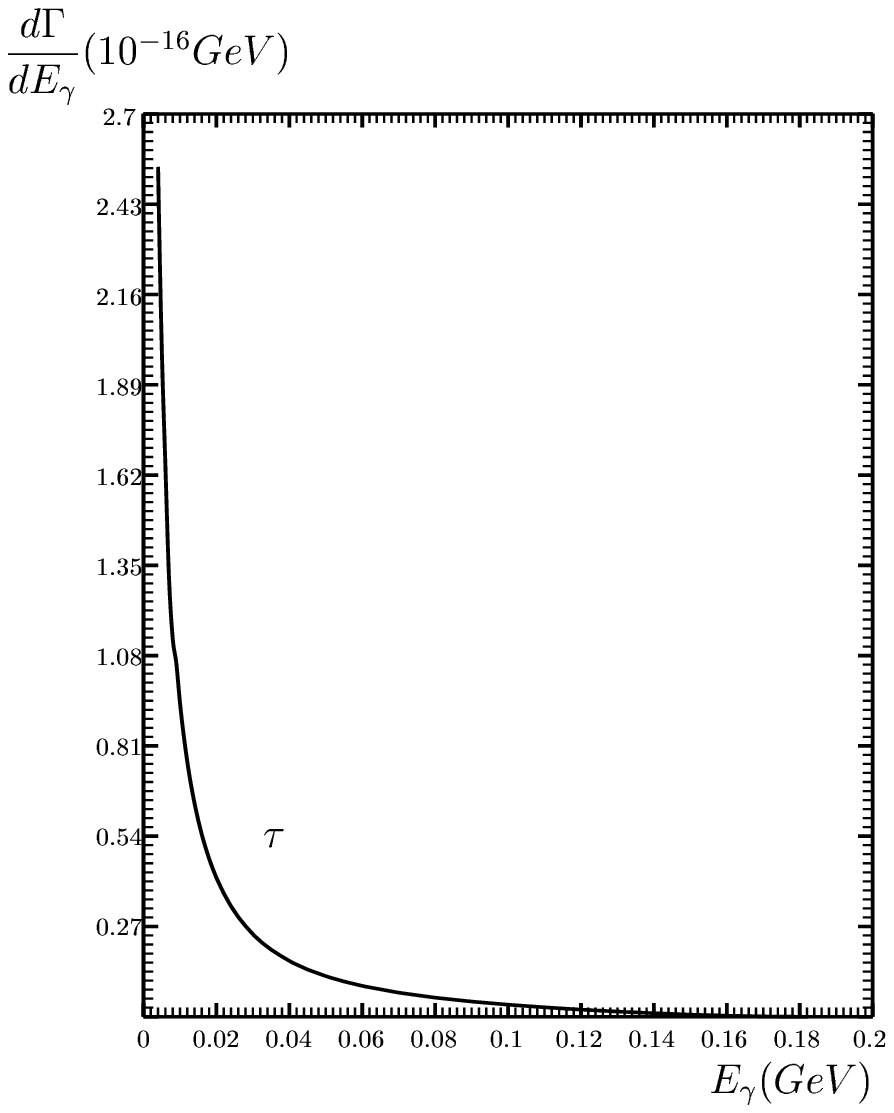, bbllx=216pt,bblly=365pt,bburx=453pt,bbury=664pt,
width=5cm,angle=0}
\caption{\bf Photon energy spectra of radiative decays 
 $D_s\longrightarrow\tau{\nu_{\tau}}
\gamma$.}
%\label{fig}
\end{center}
\end{figure}

\end{document}